# Performance Comparison of Incremental K-means and Incremental DBSCAN Algorithms

Sanjay Chakraborty
National Institute of Technology
(NIT) Raipur, CG, India.

Prof. N.K.Nagwani
National Institute of Technology
(NIT) Raipur, CG, India.

Lopamudra Dey
University of Kalyani
Kalyani,Nadia,India

## ABSTRACT
Incremental K-means and DBSCAN are two very important and popular clustering techniques for today's large dynamic databases (Data warehouses, WWW and so on) where data are changed at random fashion. The performance of the incremental K-means and the incremental DBSCAN are different with each other based on their time analysis characteristics. Both algorithms are efficient compare to their existing algorithms with respect to time, cost and effort. In this paper, the performance evaluation of incremental DBSCAN clustering algorithm is implemented and most importantly it is compared with the performance of incremental K-means clustering algorithm and it also explains the characteristics of these two algorithms based on the changes of the data in the database. This paper also explains some logical differences between these two most popular clustering algorithms. This paper uses an air pollution database as original database on which the experiment is performed.

## 1. INTRODUCTION
Clustering is a method of grouping similar types of data. This is very useful method applied in various applications. The K-means clustering and DBSCAN (Density-Based Spatial Clustering of Applications with Noise) clustering are the two most commonly used clustering techniques which are grouped the data together based on different criteria. Incremental clustering is their extended version which is suitable for the frequently change databases. Incremental K-means and DBSCAN clustering algorithms have been proposed in the papers [1, 2] and performance of incremental K-means clustering has been analysed and evaluated in paper [3] elaborately. The comparison between the typical K-means and incremental K-means has been also discussed in the paper [3].
Actual K-means suffers from several drawbacks, such as it needs predefined number of clusters and most importantly it does not has the capability to handle noisy data or outliers. Also it cannot form non-convex shapes clusters. But DBSCAN clustering is free from all these drawbacks and most importantly it can handle noisy data or outliers so efficiently. Thus these two clustering techniques are also efficiently applied on incremental databases where data are updated frequently. K-means clustering is renowned for its simplicity rather than DBSCAN clustering. In this paper K-means clustering and DBSCAN clustering are applied on a common incremental database (air pollution database) and compare their performances when the data are changed in the database. This paper also describes which clustering techniques are behaves better for % $\hat{\delta}$ (delta) changes in the original database. These algorithms and mathematical explanations of incremental K-means and DBSCAN clustering have been already proposed and discussed in the papers [1] and [2].

The rest of this paper is organized as follows. Section 2 discusses related works on these both clustering techniques. The logical comparisons between incremental K-means and DBSCAN clustering algorithms are discussed in Section 3. Section 4 describes the experimental results. Subsection of section 4 describes the experimental setup and performance evaluations of DBSCAN clustering algorithm and the performance comparison between incremental K-means and DBSCAN clustering algorithms respectively. Section 5 concludes with a summary of those clustering techniques. 'References' finally follows the 'Conclusion'.

## 2. RELATED WORK
Lot of works has been done on both K-means and DBSCAN clustering. They are also famous for their incremental nature. Sometimes they have been used together.
A paper based on the clustering of a large spatial database by the help of DBSCAN, K-means and SOM clustering algorithms is proposed. This paper analyzing the properties of density based clustering characteristics and also evaluates the efficiency of these three clustering algorithms on that particular spatial database. Finally, DBSCAN responds well to the spatial data sets [4].
A paper describes telecom churn management by comparing different clustering techniques, such as DBSCAN, K-means, EM and Farthest-First clustering techniques. In this paper DBSCAN is compared with other clustering for profiling customer segment of GSM sector. As a result, DBSCAN has seemed more suitable than K-Means, Expectation Maximization and Farthest-First for GSM operators to churn management [5].
DBSCAN and K-means clustering are suffering by several drawbacks. An approach is proposed to overcome the drawbacks of DBSCAN and K-means clustering algorithms.

This approach is known as a novel density based K-means clustering algorithm (Dbkmeans). This experiment is mainly done based on spatial data mining concept. The result will be an improved version of K-means clustering algorithm. This algorithm will perform better than DBSCAN while handling clusters of circularly distributed data points and slightly overlapped clusters. Dbkmeans is also applicable in medical data mining field [6].

Now a day clustering is used in storm detection purpose. It is a very interesting field where clustering approach is applied. In such a situation, many storms may be detected and are normally clustered corresponding to several local storms. K-means and DBSCAN clustering techniques are evaluated for their performance to cluster individual storms detected from real-time WSR-88D radar data. Based on this research, a storm clustering method is proposed that can automatically group individual storm events into a limited set of spatial clusters [7].

DBSCAN and K-means clustering are also used in network traffic classification. The analysis is based on each algorithm's ability to produce clusters that have a high predictive power of a single traffic class, and each algorithm's ability to generate a minimal number of clusters that contain the majority of the connections. In this case DBSCAN performs better than K-means clustering [8].

## 3. LOGICAL COMPARISONS

Comparison between two algorithms means compare their characteristics, their behaviour, their speed of processing and mostly their time complexities. This paper compares between the incremental behaviour of the two most popular clustering techniques (K-means and DBSCAN). The term incremental means "% of $\delta$ change in the original database" that is insertion of some new data items into the already existing clusters.

### 3.1 Cluster shapes

The first comparison lying for incremental clustering is that when new data are coming into the old database, then sometimes new clusters are formed. In case of K-means clustering the cluster shapes must be fixed means it cannot build non-convex shapes clusters. But in case of DBSCAN clustering, it discovers new clusters of arbitrary shape depends on its radius eps($\epsilon$) and Minpts(minimum number of points) discussed in paper [2]. It does not follow any fixed shape like K-means clustering. The following figure shows this difference clearly.

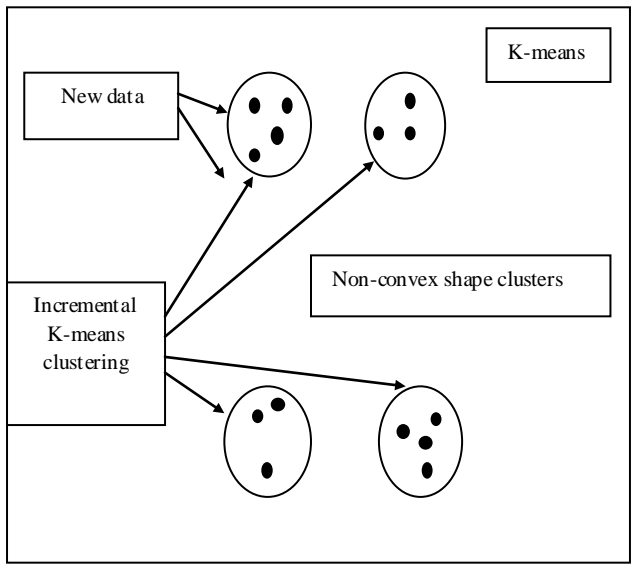

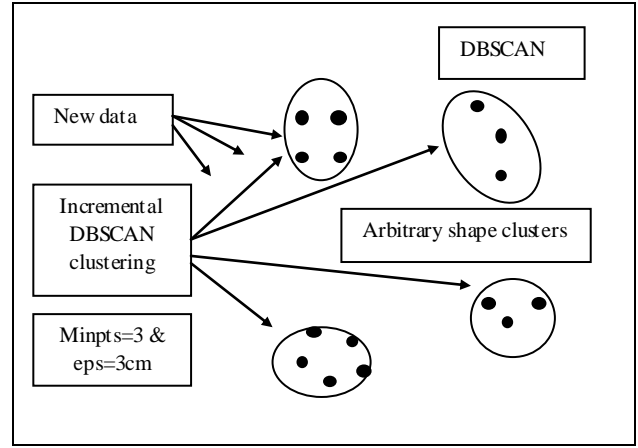

**Figure 1: Shaping difference between K-means and DBSCAN clustering respectively**

### 3.2 Predefined clusters numbers

The 'Figure.1' describes that the K-means clustering forms common shapes clusters and the DBSCAN clustering form different shapes of clusters. The second comparison lies between the concepts of these two clustering algorithms. In case of K-means clustering the total number of clusters must be predefined but in case of DBSCAN clustering the clusters are formed based on the new coming data, there is no need to predefine the number of clusters.

### 3.3 Outliers handling

But the main comparison lies between these two on the basis of handling noisy data or outliers. This is very important task of handling noisy data properly when building clusters on large dynamic database. The K-means clustering algorithm is sensitive to noise and outlier data points because a small number of such data can substantially influence the mean value. But the DBSCAN algorithm has the ability to efficiently handle the noisy data even in the dynamic environment where the data are changed randomly. The following example shows this mathematically,

*Example.1*

Suppose there are nine data in a database, such as (4,6), (112,94), (9,15), (4,9), (8,17), (3,2), (1,4), (1,7) and (10,9). First K-means clustering is applied after assuming total number of cluster K=3 and means are (4,6), (4,9) and (3,2) respectively. So, if Manhattan distance function is used then,

Cluster 1= |(9-4)+(15-6)|=|5+9|=14
Cluster 2= |(9-4)+(15-9)|=|5+6|=11 (minimum)
Cluster 3= |(9-3)+(15-2)|=|6+13|=19

Thus the data (9,15) should be entered into cluster 2. In the same way other data of the database are clustered properly except the data (112, 94). So, this data are treated as outliers or noisy data. K-means clustering is unable to handle such noisy data. In case of dynamic environment, when the new data are inserted into the existing database, then in the incremental approach they are directly clustered those data after comparing them from the means of the existing clusters. This concept is clearly discussed in the paper [1].

So, if two new coming data such as (155,112) and (99,125) which are also out of range just like the previous outlier, then those new coming data are not handled by the K-means clustering. The following figure shows the approach of handling noisy data by the K-means clustering algorithm clearly. It describes that the three data (112, 94), (155,112) and (99,125) are outliers and they cannot enter into any clusters.

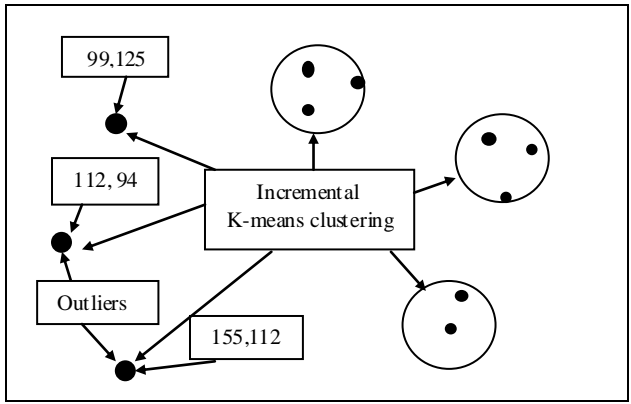

**Figure 2: Handling outliers by incremental K-means clustering**

But in the same case if incremental DBSCAN clustering is applied then it follows the same principal of clustered the new incremented data just like incremental K-means clustering except that it is able to handle the noisy data or outliers properly. As per the above example if those two new coming data (155,112) and (99,125) are entered into the old database, then incremental DBSCAN clustered those two new noisy data with the previous noisy data (112, 94) only if they satisfy the Minpts and eps conditions [mean-distance<=eps & size(cluster)>Minpts].

The following figure shows the concept of handling outlier by the incremental DBSCAN clustering algorithm. So, cluster 4 is built by three noisy data, such as
Cluster 4= [(112, 94), (155,112), (99,125)].

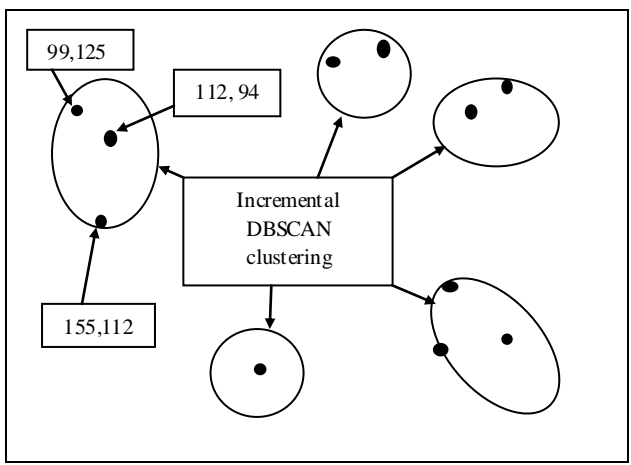

**Figure 3: Handling outliers by incremental DBSCAN clustering**

Due to this noisy data handling capability incremental DBSCAN requires more processing time compare to incremental K-means clustering. The performance evaluation section shows this fact clearly.

## 4. EXPERIMENTAL RESULTS
This paper implements the performance comparison between the incremental K-means and the incremental DBSCAN clustering algorithms. The required experimental setup for doing this performance comparison is described below,

### 4.1 Experimental Setup
This experiment is done on air-pollution database with the help of Java language, Weka interface and other tools.
This analysis is based on the observation of the air pollution data has been collected from "West Bengal Air Pollution Control Board". This database consists of four air-pollution elements or attributes. In this paper both the algorithms are developed in Java 1.5. Weka (Waikato Environment for Knowledge Analysis) is the other open source API's (Application Programming Interfaces) to support the other functionalities. Weka is used for performing some data mining related operations. Eclipse is used as a development IDE (Integrated Development Environment) for java and library of other technologies are added as external jar (Java Archives) in the eclipse. Finally, Mysql is used to construct databases.
All the experiments are performed on a 2.26 GHz Core i3 processor computer with 4GB memory, running on Windows 7 home basic.

### 4.2 Performance Evaluations
The performance evaluation of the incremental K-means clustering algorithm has been already developed and discussed elaborately in the paper [3]. In this paper, the performance of the incremental DBSCAN clustering algorithm is evaluated. This algorithm has been already proposed in the paper [2]. Here, it is observed that how the incremental DBSCAN algorithm behaves when new data are inserted into the existing database. Here, the changing time is measured with the change of the data in the original database. This paper also discusses the performance comparison between these two incremental clustering algorithms.
To evaluate the performance of incremental DBSCAN algorithm, first calculate the change of time (milliseconds) with the increment of data in the original database. This increment of data is known as %delta change in the original database. The following table and figure explains that how the existing DBSCAN clustering works with the change of data in the database.

**Table.1 Time vs. data in actual DBSCAN clustering**

| Original Data | Time (ms) |
|---|---|
| 500 | 40,250 |
| 600 | 41,500 |
| 700 | 43,300 |
| 800 | 48,230 |
| 900 | 50,720 |
| 1000 | 52,324 |
| 1100 | 53,460 |
| ……. | ……. |

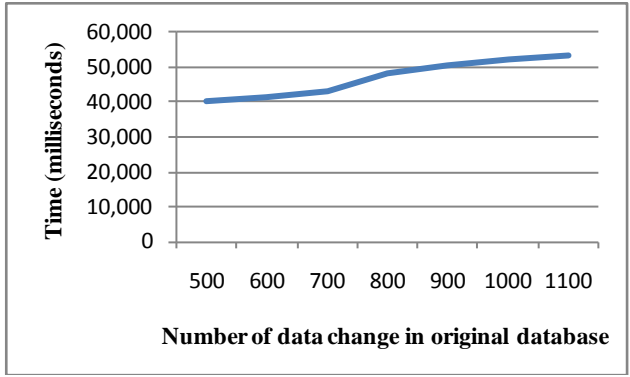

**Figure 4: Graph for actual DBSCAN result**

Figure 4 describes that how the time slowly increases with the increases of data in the original database. Now when the new data are inserted into the old database, then for that new data the proposed incremental DBSCAN clustering algorithm is applied. This algorithm directly clustered the new coming data without rerunning the DBSCAN algorithm by comparing those data with the means of existing clusters.

**Table 2. Time vs. incremented data in incremental DBSCAN clustering**

| Incremental Data | Time (ms) |
|---|---|
| 100 | 12,480 |
| 200 | 24,643 |
| 300 | 38,943 |
| 400 | 52,530 |
| 500 | 60,930 |
| ……. | ……. |

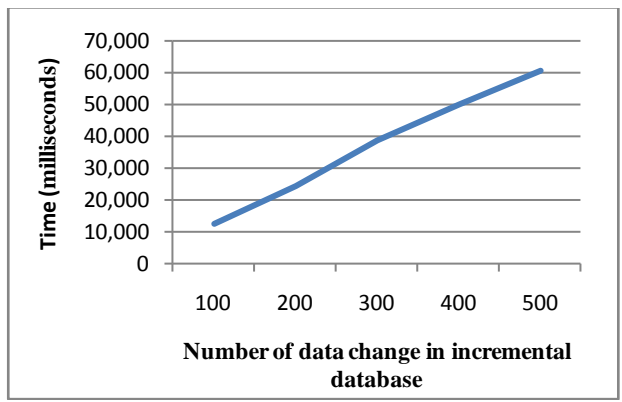

**Figure 5: Graph for incremental DBSCAN result**

Figure 5 describes how the time rapidly increases with the increases of data in the incremental database. Now, it can easily calculate after combining the above two results that for what % of delta ($\delta$) change [insertion of some new data items into the already existing database] in the database up to which the incremental DBSCAN clustering behaves better than the actual DBSCAN clustering. First calculate all the delta changes of this database by the help of following formula.

$$\% \, \delta \text{ change in DB} = \frac{(NEW\ DATA - OLD\ DATA)}{OLD\ DATA} \times 100 \quad (1)$$

**Table 3. Time vs. %$\delta$ change in DB for both actual and incremental DBSCAN clustering**

| Actual Time(ms) | %$\delta$ change in the database | Incremental Time(ms) |
|---|---|---|
| 41,500 | $\delta_1 = \frac{(600-500)}{500} \times 100 = 20\%$ | 12,480 |
| 43,300 | $\delta_2 = 40\%$ | 24,643 |
| 48,230 | $\delta_3 = 60\%$ | 38,943 |
| 50,720 | $\delta_4 = 80\%$ | 52,530 |
| …….. | ………. | ……. |

From the above calculation the particular threshold value upto which the proposed DBSCAN clustering behaves better than the existing one is 72% (Cut-off point). The following figure shows it clearly.

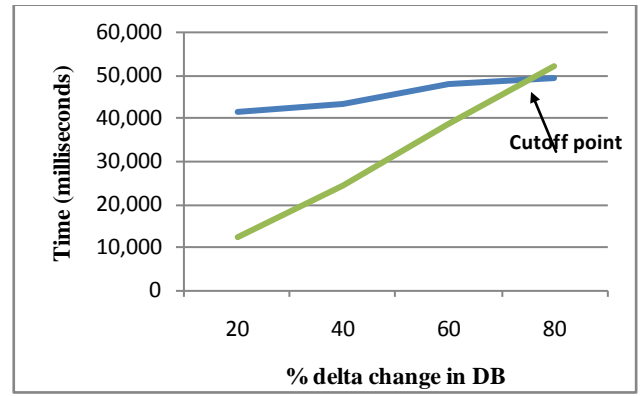

**Figure 6: Graph for actual DBSCAN vs. incremental DBSCAN**

### 4.3 Performance comparison

Now, performance evaluation of incremental DBSCAN and incremental K-means can be compared easily. This comparison is based on the logic of that for every % of delta change in the database how the incremental K-means and DBSCAN algorithms are performed different from each other. All the experiments are performed on the air-pollution database explain in the paper [3]. The following figures describe it clearly.

From the below figures, it can be easily understood that the incremental K-means clustering algorithm is better than the incremental DBSCAN clustering because incremental K-means takes less amount of time for the particular change of data in the database whereas incremental DBSCAN takes much larger amount of time. DBSCAN takes more time because it requires extra time to properly handle and clustered the noisy data but K-means never waist time to handle those outliers.

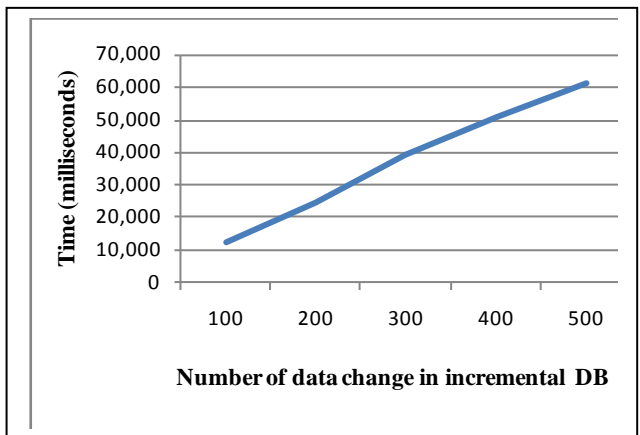

**Figure 7(A): Graph shows the performance of incremental DBSCAN clustering**

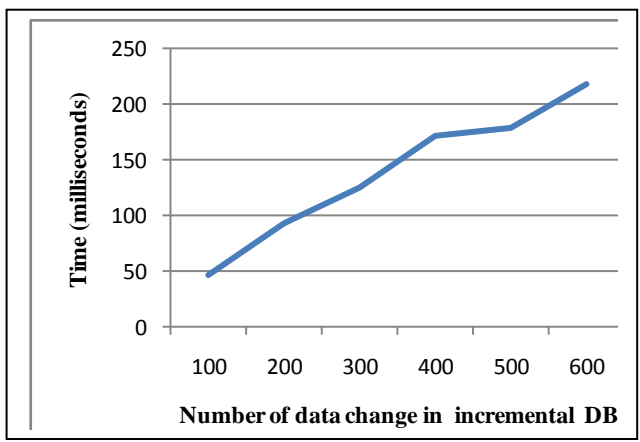

**Figure 7(B): Graph shows the performance of incremental K-means Clustering**

The graph of 'Figure.7(B)' has been collected from the paper [3].

## 5. CONCLUSION
In this paper the performance evaluation of a proposed incremental DBSCAN clustering algorithm is established. This paper also logically compares the characteristics of incremental DBSCAN and incremental K-means clustering algorithms. It also compares the performance of these two algorithms when they are applied on real time dynamic databases. As a result, the incremental K-means clustering performs better than the incremental DBSCAN clustering with respect to time analysis.

## 6. ACKNOWLEDGMENT
Special thanks to Dr. S. Verma and Dr. T.S. Sinha whose comments improved the presentation of this article.